\documentclass[final,5p,times,twocolumn,sort&compress]{elsarticle}
\pdfoutput=1
\usepackage[hyphens]{url}
\usepackage{amsmath}
\usepackage{graphicx}
\usepackage{amssymb}
\usepackage{units}
\usepackage[hyperindex,breaklinks]{hyperref} 
\DeclareGraphicsExtensions{.pdf}

\usepackage{preprintcover}  % load package
\PreprintCoverPaperTitle{Search for long-lived, multi-charged particles in $pp$ collisions at $\sqrt{s}=7$~TeV using the ATLAS detector}  % paper title
\PreprintIdNumber{CERN-PH-EP-2012-360}  % CERN preprint number
\PreprintCoverAbstract{A search for highly ionising, penetrating particles with electric charges from $|q| = 2e$ to $6e$
is performed using the ATLAS detector at the CERN Large Hadron Collider.
Proton-proton collision data taken at $\sqrt{s}=7$~TeV during the 2011 running period,
corresponding to an integrated luminosity of \intlumi, are analysed. 
No signal candidates are observed, and 95\% confidence level cross-section upper limits are interpreted as mass-exclusion lower limits for a simplified Drell--Yan production
model. In this model, masses are excluded from \unit[50]{\GeV{}} up to 430, 480, 490, 470 and \unit[420]{\GeV{}} for charges $2e$,
$3e$, $4e$, $5e$ and $6e$, respectively. }  % paper abstract
\PreprintJournalName{PLB}  % journal name
\newcommand{\dedx}{\ensuremath{\textrm{d}E/\textrm{d}x}}

\newcommand{\ndata}{\ensuremath{N_{\textrm{data}}}}
\newcommand{\nrec}{\ensuremath{N^{\textrm{rec}}_\textrm{data}}}

\newcommand{\qballs}{\ensuremath{Q\textrm{-balls}} }

\newcommand{\beq}{\begin{equation}}
\newcommand{\eeq}{\end{equation}}
\newcommand{\bea}{\begin{eqnarray}}
\newcommand{\eea}{\end{eqnarray}}
\newcommand{\intlumi}{$4.4$~fb$^{-1}$}
\newcommand{\htCut}{0.4}
\newcommand{\pixSCut}{10}

\newcommand{\backgroundEstQTwoValue}{\ensuremath{0.41}}
\newcommand{\backgroundEstQThreeValue}{\ensuremath{1.37}}
\newcommand{\backgroundEstQTwoError}{\ensuremath{0.08}}
\newcommand{\backgroundEstQTwoSys}{\ensuremath{0.02}}
\newcommand{\backgroundEstQThreeError}{\ensuremath{0.46}}
\newcommand{\backgroundEstQThreeSys}{\ensuremath{0.07}}
\newcommand{\backgroundEstQTwo}{\backgroundEstQTwoValue\ensuremath{\pm}\backgroundEstQTwoError}
\newcommand{\backgroundEstQThree}{\backgroundEstQThreeValue\ensuremath{\pm}\backgroundEstQThreeError}
\newcommand{\backgroundEstQTwoStatSys}{\backgroundEstQTwoValue\ensuremath{\pm}\backgroundEstQTwoError\ensuremath{\pm}\backgroundEstQTwoSys}
\newcommand{\backgroundEstQThreeStatSys}{\backgroundEstQThreeValue\ensuremath{\pm}\backgroundEstQThreeError\ensuremath{\pm}\backgroundEstQThreeSys}
\def\pt{\ensuremath{p_{\textrm{T}}}} % Subscript roman not italic (EE)
\def\TeV{\ifmmode {\mathrm{\ Te\kern -0.1em V}}\else
                   \textrm{Te\kern -0.1em V}\fi}%
\def\GeV{\ifmmode {\mathrm{\ Ge\kern -0.1em V}}\else
                   \textrm{Ge\kern -0.1em V}\fi}%
\def\et{\ensuremath{E_{\textrm{T}}}} % Subscript roman not italic (EE)

\begin{document}

\begin{frontmatter}

\title{Search for long-lived, multi-charged particles in $pp$ collisions at $\sqrt{s}=7$~TeV using the ATLAS detector}
\author{The ATLAS Collaboration}
%\maketitle

\begin{abstract}
A search for highly ionising, penetrating particles with electric charges from $|q| = 2e$ to $6e$
is performed using the ATLAS detector at the CERN Large Hadron Collider.
Proton-proton collision data taken at $\sqrt{s}=7$~TeV during the 2011 running period,
corresponding to an integrated luminosity of \intlumi, are analysed. 
No signal candidates are observed, and 95\% confidence level cross-section upper limits are interpreted as mass-exclusion lower limits for a simplified Drell--Yan production
model. In this model, masses are excluded from \unit[50]{\GeV{}} up to 430, 480, 490, 470 and \unit[420]{\GeV{}} for charges $2e$,
$3e$, $4e$, $5e$ and $6e$, respectively.

\end{abstract}


\end{frontmatter}

\section{Introduction}
Numerous theories of physics beyond the Standard Model (SM) predict long-lived\footnote{The term long-lived in this paper refers to a
  particle that does not decay within the ATLAS detector.} exotic 
objects producing anomalous ionisation. These include magnetic monopoles
\cite{monopoles}, dyons \cite{dyons}, 
long-lived  micro black holes in models of low-scale gravity \cite{bh} 
and \qballs \cite{Qballs}, which are non-topological solitons predicted by minimal
supersymmetric generalisations of
the SM.
No such particles have so far been observed in cosmic-ray and collider searches \cite{monopoles,SLIM,Abazov:2008qu,Aaltonen:2009kea}, including several recent 
searches at the Large Hadron Collider (LHC) \cite{HIP2010,SMP2010,SMP2012,Khachatryan:2011ts,Chatrchyan:2012sp,monopole2012}. %CMS-PAS-EXO-11-090
The high centre-of-mass energy of the LHC makes a new energy regime accessible, and searching for
 multi-charged particles with  electric charges $2e \le
|q| \le 6e$ complements the searches for slow singly charged particles \cite{SMP2012}
 and for particles with charges beyond $6e$ \cite{HIP2010}.

The existence of long-lived particles with an electric charge $|q| > e$ could have implications for the formation of composite dark matter \cite{Khlopov:2011tn}. 
Two extensions of the SM in which heavy
stable multi-charged particles are predicted are the AC model \cite{5} and the walking technicolour
model \cite{Sannino,Gudnason,Foadi}. The AC model is  based on the approach
of almost-commutative geometry \cite{bookAC} which extends the fermion content of the SM
 by two heavy particles with opposite electric charges, $\pm q$. %where $|q|$ may be larger than $e$.
The minimal walking technicolour model predicts the existence of three particle pairs, with electric charges given in general by $q+e$, $q$, and $q-e$, which
would behave like leptons in the detector.
In both of these models, $|q|$ may be larger than $e$.

This Letter describes a search for multi-charged particles  in $\sqrt{s}=7$~TeV
$pp$ collisions using data collected in 2011 by the ATLAS detector at the CERN LHC. 
The data sample corresponds to an integrated
luminosity of \intlumi.
Multi-charged particles will be highly ionising, and thus leave an abnormally large specific ionisation signal, \dedx{}.
In this Letter, a search for such particles traversing the ATLAS detector
leaving a track in the inner tracking detector, and producing a signal in the muon spectrometer, is reported.
A SM-like coupling proportional to the electric charge is assumed as the production model of the multi-charged particles. Therefore, the main production mode is Drell--Yan (DY) with no weak coupling.
Multi-charged particles can also be pair-produced from radiated photons resulting in a larger production cross section, 
and in some cases non-perturbative effects \cite{Sakharov:1948yq} can also enhance the production rate. 
In the derivation of limits, neither enhancement is included in the calculation resulting in conservative limits in these scenarios.

\section{ATLAS detector}
\label{sec:signalsimulation}

The ATLAS detector \cite{atlas:2008} covers nearly the entire
solid angle around the collision point. It consists of an inner
tracking detector (ID) comprising a silicon pixel detector (pixel), a silicon
microstrip detector (SCT) and a Transition Radiation Tracker
(TRT).  Apart from being a straw-based tracking
detector, the TRT (covering $|\eta|<2.0$)\footnote{The ATLAS coordinate system is right-handed with the pseudorapidity, $\eta$,
defined as $\eta = -\ln[\tan(\theta/2)]$, where the polar angle $\theta$ is measured with respect to the 
LHC beamline.  The azimuthal angle, $\phi$, is measured with respect to the $x$-axis, which points
towards the centre of the LHC ring.  The $z$-axis is parallel to the anti-clockwise beam viewed from
above.  Transverse momentum and energy are defined as $\pt = p \sin \theta$ and $\et = E \sin \theta$, 
respectively.} also provides particle identification via
transition radiation and ionisation energy loss measurements \cite{TRTPID}. The ID is
surrounded by a thin superconducting solenoid providing a \unit[2]{T}
axial magnetic
field, and by high-granularity liquid-argon (LAr) sampling electromagnetic calorimeters.
An iron--scintillator tile calorimeter
provides hadronic energy measurements in the central rapidity region.
The endcap and forward regions are instrumented with
LAr calorimeters for both electromagnetic and hadronic energy
measurements. 
The calorimeter system is surrounded
by a muon spectrometer (MS) incorporating three superconducting toroid
magnet assemblies. The MS is a combination of several sub-detectors used 
to measure muons that traverse the ATLAS calorimeters.
The Resistive Plate Chambers (RPC) in the barrel region ($|\eta| < 1.05$) and
the Thin Gap Chambers (TGC) in the endcap region ($1.05 < |\eta| < 2.4$) provide signals for the trigger
for charged particles reaching the MS. 
Monitored Drift Tube (MDT) chambers measure the momentum and
track positions of muons with very high precision.

\section{Simulated samples}
\label{sec:simulation}

Benchmark samples of simulated events with multi-charged particles are produced  for  masses of 50, 100, 200, 300, 400, 500 and
\unit[600]{\GeV{}}, with charges\footnote{Wherever a charge is quoted for the exotic particles, the charge conjugate state is also implied.} $2e$, $3e$, $4e$, $5e$ and $6e$. 
Pairs of long-lived multi-charged particles are simulated using {\sc MadGraph5}~\cite{madgraph5} via the DY process to model the kinematic distributions.
The DY production model also determines the cross section used for limit setting.  
Typical values for the cross sections of simulated multi-charge pair production range from tens of pb for a mass of \unit[50]{\GeV} down to a few fb at a mass of \unit[600]{\GeV}.
Events are generated using the {\sc CTEQ6L1}~\cite{Pumplin:2002vw} parton distribution functions, and {\sc Pythia} version 6.425 \cite{Sjostrand:2006za} is
used for hadronisation and underlying-event generation. 
A {\sc Geant4} simulation \cite{Agostinelli:2002hh, Aad:2010ah}
 is used to model the response of the ATLAS detector, and the samples are reconstructed and analysed in the same way
as the data.
The production cross sections are estimated using {\sc
  MadGraph5} and are cross-checked with 
{\sc CalcHEP} 3.4~\cite{Belyaev:2012qa}. 
Each simulated event is overlaid with additional collision events
(``pile-up'') in order to reproduce the observed 
distribution of the number of proton--proton collisions per bunch crossing.
In 2011 data the average number of interactions per bunch crossing was typically between 5 and 20.”
These samples are used to determine the
detection efficiency, the resolution on the quantities used in the event selection and the associated
systematic uncertainties for multi-charged particles.   
While the background estimation is data-driven, muons from $Z\to\mu\mu$ simulated samples are  used to calibrate
 the selection variables. 
These samples are generated in {\sc Pythia} and passed through the {\sc Geant4}  simulation of the ATLAS detector.

\section{Ionisation estimators}

The specific energy loss, \dedx{}, is described by the Bethe--Bloch
formula \cite{bethe}. The energy loss depends quadratically on the particle charge,
$q$, so that  particles with higher charges have
a significantly higher energy loss.

\subsection{MDT \dedx{} }
Each drift tube of the MDT system provides a signal proportional to 
the charge from ionisation, which is used to estimate \dedx{}.
A truncated mean of \dedx{}, where the maximum value is removed, is used as the overall MDT \dedx{} estimator.
As each track crosses more than 20 drift tubes, the MDT \dedx{} provides a good estimate of ionisation losses.

\subsection{TRT \dedx{} }

Energy deposits in a TRT straw greater than \unit[200]{eV} (low-threshold
hits) are used for tracking, while those that exceed \unit[6]{keV}
(high-threshold hits) occur due to the passage of highly ionising particles or due to transition
radiation emitted by highly relativistic electrons when they
cross radiator material between the straws.
The estimated \dedx{} value for each hit is derived from the time the signal
remains above the low threshold.   
The TRT \dedx{} is the truncated mean of the \dedx{} estimates, where
the highest estimate is removed. %of the hits.  
On average, a track in the TRT contains 32 hits. %The number of straws exceeding the high-threshold (HT) provides a second measurement of
Additionally, the ratio of the number of high-threshold (HT) hits to the total number of TRT 
hits on a given track $f^{\textrm{HT}}$ provides a second estimator of high ionisation.

\subsection{Pixel \dedx{} }

The pixel detector %provides direct information about 
measures the 
charge from ionisation in each pixel. 
The  \dedx{} from the pixel detector is calculated from the
truncated mean of measurements from several clusters of pixels \cite{pixeldedxnote}.
Particles with charges higher than $2e$ deposit energies which easily exceed the dynamic range of the pixel detector readout. Therefore, the electronic signal is saturated and pixel information will not be read out  leading to an unreliable \dedx{} measurement for such particles.

\subsection{\dedx{}  significance}

The significance of each \dedx{} variable is defined as the difference between the
observed \dedx{} of the track and that expected for muons,
measured in units of the uncertainty of the measurement:

\begin{equation}
S(\dedx{}) = \frac{\text{\dedx{}}_{\text{track}} - \langle\text{\dedx{}}_{\mu}\rangle}{\sigma(\text{\dedx{}}_{\mu})}.
\end{equation}
Here $\text{\dedx{}}_{\text{track}}$ represents the estimated \dedx{} of the track, 
and $\langle\text{\dedx{}}_{\mu}\rangle$ and $\sigma(\text{\dedx{}}_{\mu})$, respectively, are the mean and the width
of the \dedx{} distribution for muons in data.

\begin{figure}[h!]
  \begin{center}
    \includegraphics[width=0.45\textwidth]{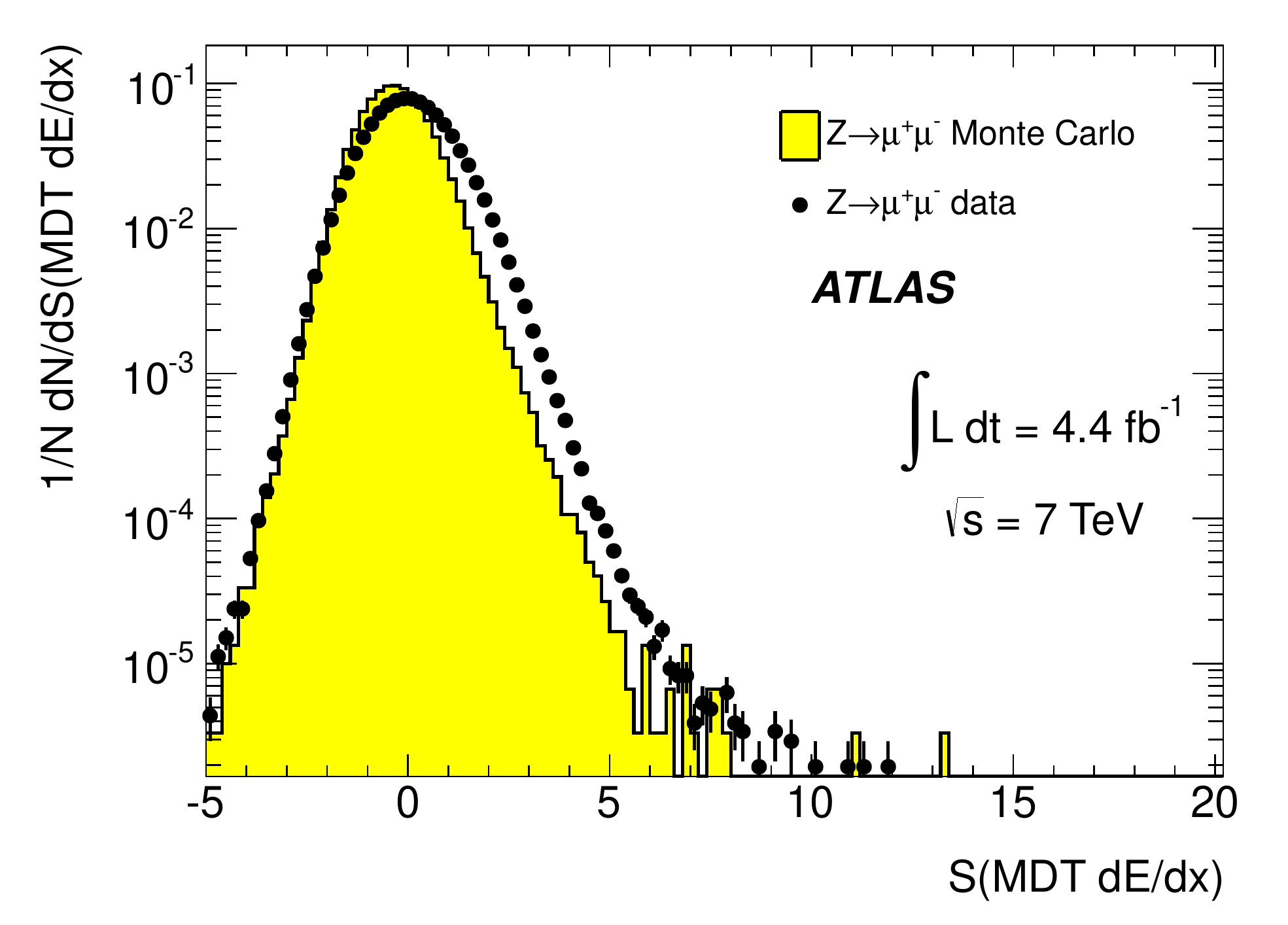}
    \includegraphics[width=0.45\textwidth]{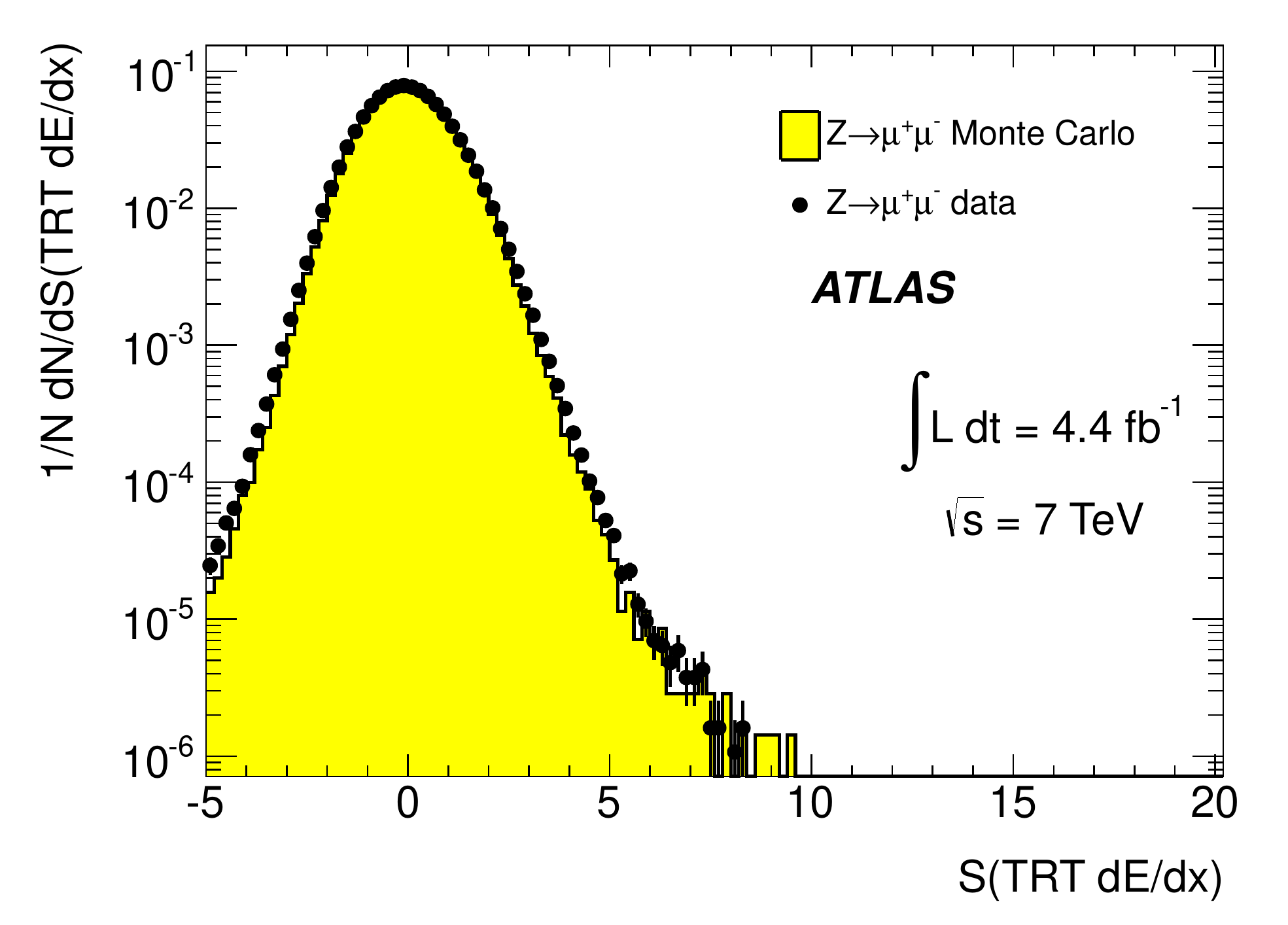}
  \end{center}
  \caption{Comparison of normalised distributions of the $S$(MDT \dedx{}) (top) and $S$(TRT \dedx{})  (bottom) 
    for muons from $Z\rightarrow \mu\mu$ events in data and simulation. } 
  \label{fig:dataMCagreement}
\end{figure}

\begin{figure}[h!]
  \begin{center}
    \includegraphics[width=0.45\textwidth]{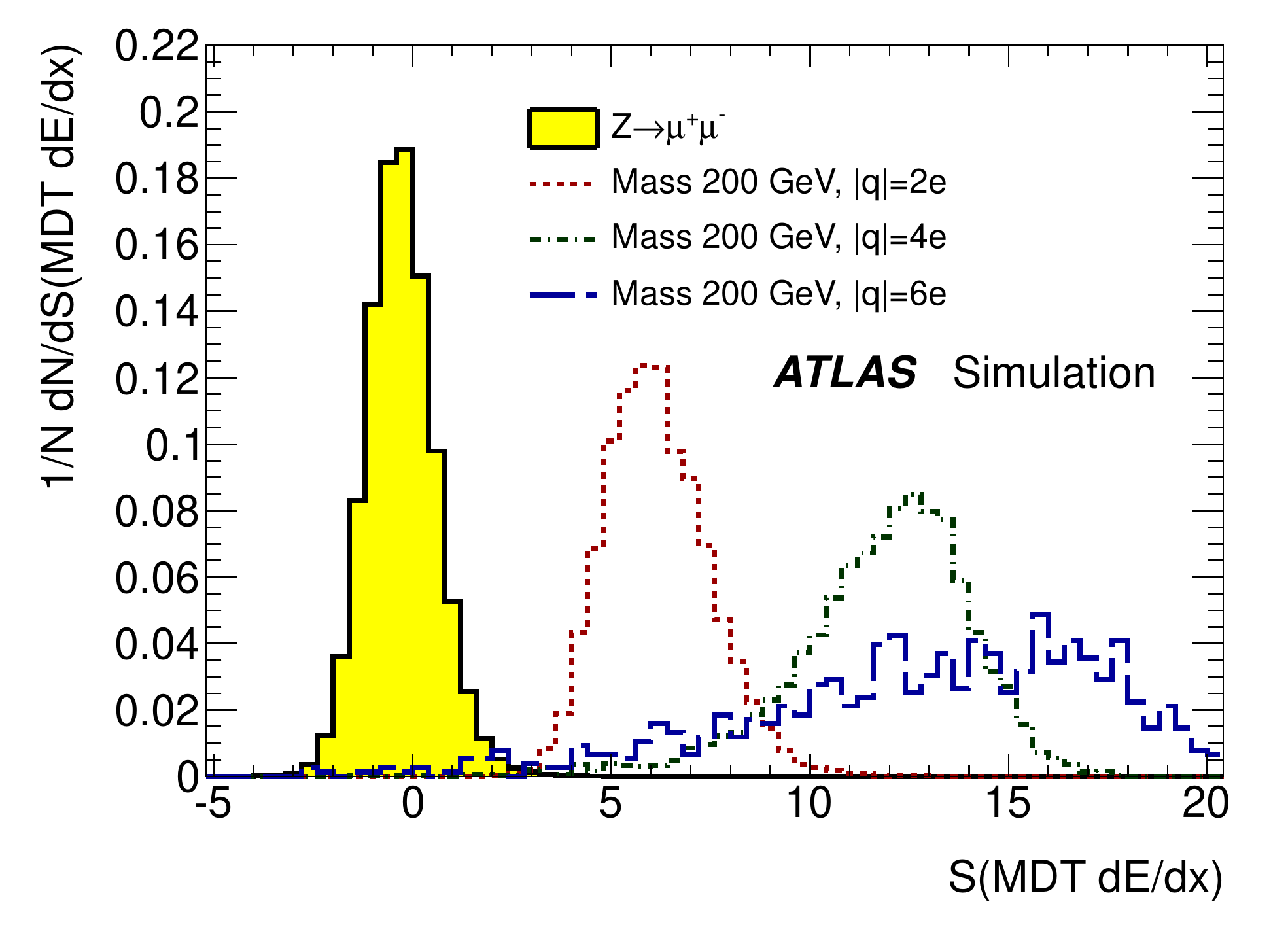}
    \includegraphics[width=0.45\textwidth]{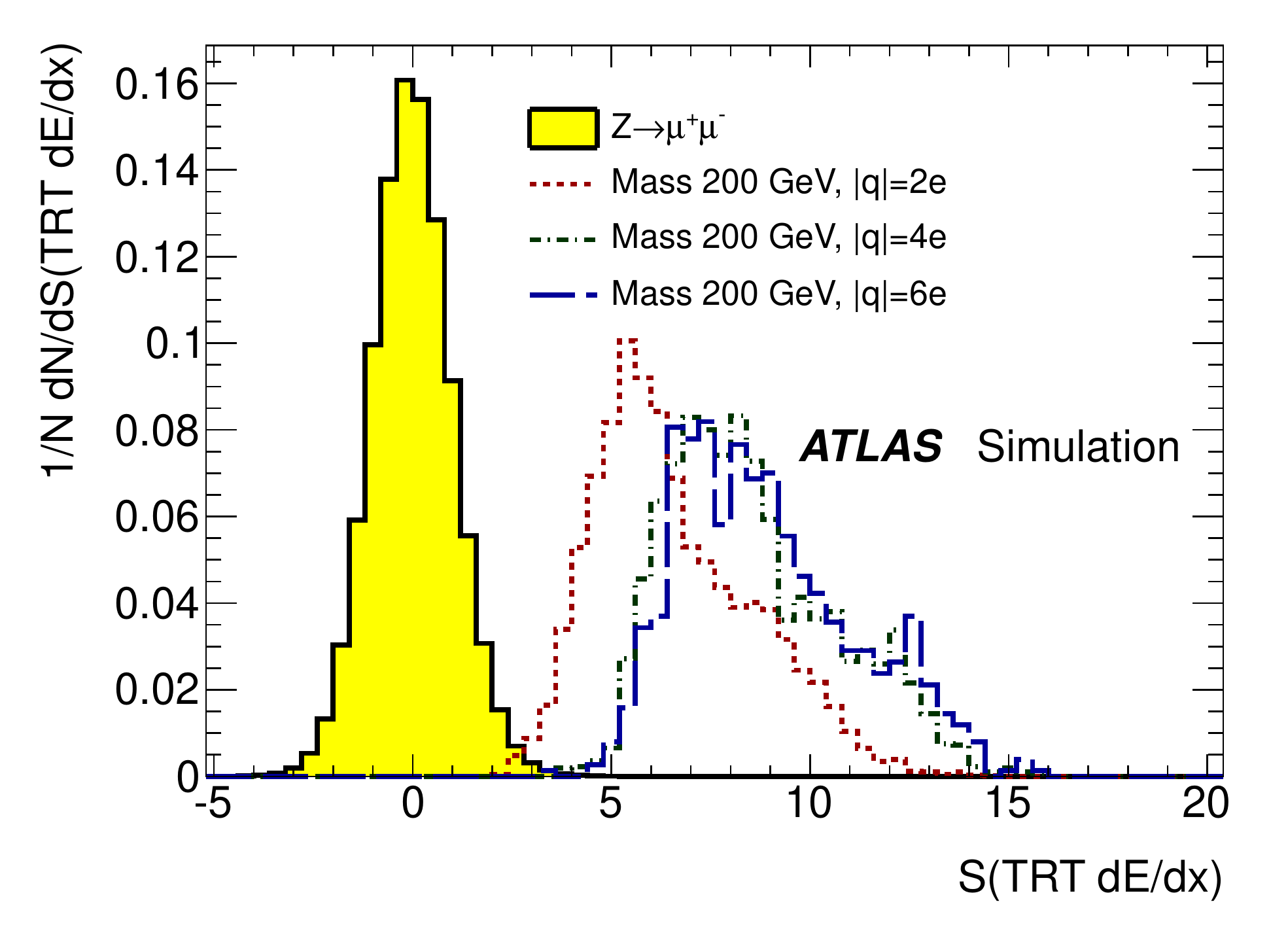}
  \end{center}
  \caption{Normalised distributions of $S$(MDT \dedx{}) (top) and $S$(TRT \dedx{}) (bottom) for simulated muons and multi-charged particles. %in arbitrary units. 
    Distributions are shown for the signal samples for $|q|$ = $2e$, $4e$ and $6e$, for a mass of \unit[200]{\GeV{}}.
}
  \label{fig:TRTseparation}
\end{figure}

To obtain expected \dedx{} values and their resolution for the
different detector components (MDT, TRT, Pixel), the \dedx{}  variables are calibrated with muons 
from $Z\to\mu\mu$ events in data and simulation.
Muons  for this calibration are selected by requiring a track reconstructed in the MS matched to a good quality  track in the ID with 
$p_{\text{T}} >$ \unit[20]{\GeV{}} and $|\eta| <$ 2.4. Each muon is further
required to belong to an oppositely charged pair with dimuon mass between \unit[81]{\GeV{}} and  \unit[101]{\GeV{}}.
Fig.~\ref{fig:dataMCagreement} shows the comparison between these muons in data and simulation for the MDT and TRT \dedx{} significance.
While the TRT distribution shows good agreement except in the tails, a discrepancy between simulation and data is observed for the MDT significance.
This discrepancy has a small effect on the limit setting, and the effect is included in the systematic uncertainties.
Fig.
~\ref{fig:TRTseparation} shows the distributions of the MDT and TRT \dedx{} significance for simulated muons from 
$Z\to\mu\mu$ production compared to those of multi-charged particles
for different charges ($2e$, $4e$ and $6e$) and for a mass of \unit[200]{GeV}. 
For the multi-charge particle search, the $S$(MDT \dedx{}) and
      $S$(TRT \dedx{}) variables are required to exceed threshold
      values. These thresholds are established from the separation of
      the \dedx{} significance distributions between muons and $|q| = 2e$
      signal particles. The \dedx{} significance distributions for
      higher charge values, $|q| > 2e$, are further separated from muons,
      as seen for simulated events in Fig.~\ref{fig:TRTseparation}. The detailed
      response for these higher charge particles may not be perfectly
      modelled by the simulation due to saturation effects.  However, their
      \dedx{} response will certainly be higher than that of $|q| = 2e$
      particles, and thus their detailed response has no significance
      for the analysis.
The separation power of the pixel \dedx{} significance is shown in Fig. \ref{fig:pixelseparation} for a $2e$ charge at $m = $ 200, 400 and 600 \GeV{}. 
The behaviour of the \dedx{} significance distributions is found to be as expected with respect to  $p_{\text{T}}$, $\eta$, and $\phi$.
For simulated multi-charged particles the \dedx{} significances strongly depend on the particle's charge and weakly on the particle's mass.

\section{Event and candidate selection}
\label{sec:selection}

Multi-charged candidates are sought for among those particles traversing the entire ATLAS detector, thus being initially selected as muons.
Candidates are selected by analysing the specific ionisation losses in the different detectors.   
The search is based on a cut-and-count
method, described in Section~\ref{sec:abcd}, where the signal region is defined by  high \dedx{}
significances of the track measured  by the TRT and MDT detectors. 
  
Track reconstruction assumes
particles with charge $\pm1e$, whereas particles with higher charges
bend more in the magnetic field.
Therefore, the effective cut on the momentum of the multi-charged particle imposed by the trigger and selection is a factor of $|q|/e$ higher 
than the cut on the muon candidate. In the following, we will refer to
$p_{\text{T}}$ as the reconstructed transverse momentum assuming
charge $|q|=1e$.

\subsection{Trigger and event selection}
Events collected with a single-muon trigger \cite{MuonPerformance}
with a transverse momentum threshold of $p_{\text{T}}= $ \unit[18]{\GeV{}} are considered.
In simulated events the trigger efficiency from the RPC is corrected as a function of a particle's $\eta$ and $\beta$, where $\beta$ is the
ratio of the particle's velocity to the speed of light. 
Events are further required\footnote{Information on the MDT \dedx{} is not available in the
standard ATLAS data stream. Hence, this analysis is based on a special stream which includes this information. The
 $p_{\text{T}}$ requirements on muons given here are imposed for the preparation of this stream and are not optimised for the
current analysis.} to contain either at least one muon with
$p_{\text{T}}>$ \unit[75]{\GeV{}} or at least two muons with $p_{\text{T}}>$ \unit[15]{\GeV{}}.

\subsection{Candidate selection}
\label{sec:cansel}

Candidate particles are tracks reconstructed in the MS which are required to be matched to the object passing the muon trigger, and to originate within tolerances from the primary interaction point. They must also be within the acceptance region $|\eta| < 2.0$, have a $p_{\text{T}} >$ \unit[20]{\GeV{}}, and leave a high-quality track in the ID. However, because of potential pixel readout saturation, there is no requirement that a candidate particle has pixel information. The $p_{\text{T}}$ measured by the muon system is smaller than the $p_{\text{T}}$ measured in the ID due to energy loss in the calorimeters, and the $p_{\text{T}}$ in the ID is used for candidate selection.  
In the track candidate selection, the measurement of the ionisation energy loss in the calorimeter system was not used.   
However, the calorimeter energy loss was validated for use as an independent cross-check in case of an observation of candidates above the expected background.

\begin{figure}[h!]
  \begin{center}   
    \includegraphics[width=0.45\textwidth]{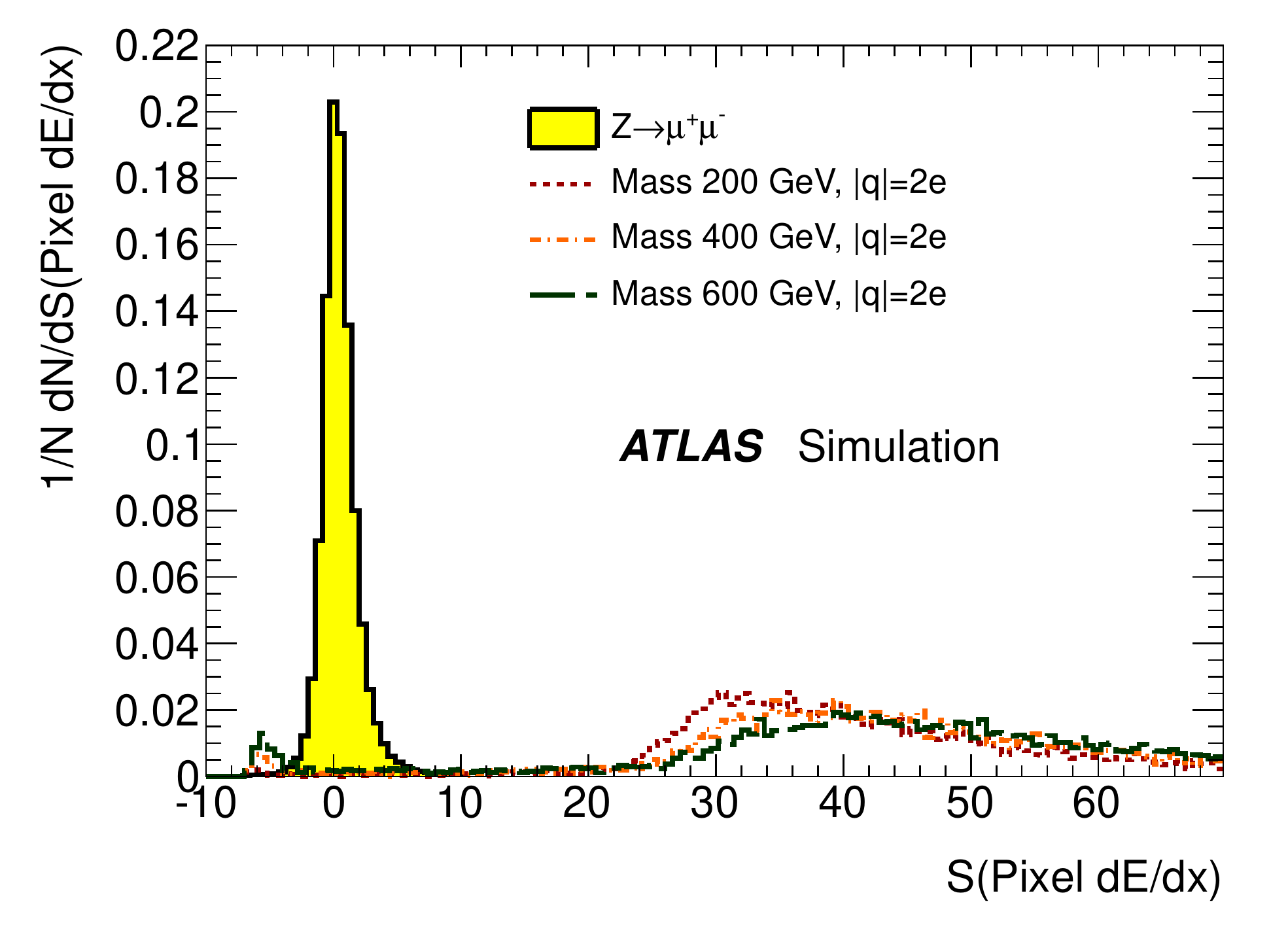}
  \end{center}
  \caption{Normalised distribution of $S$(pixel \dedx{}) for simulated muons
    and multi-charged particles. %in arbitrary units. 
    Distributions are shown for the signal sample for $|q|$ = $2e$, for masses of 200, 400 and \unit[600]{\GeV{}}. 
    The structure at a significance of -5 is from pixel read-out saturation.
} 
  \label{fig:pixelseparation}
\end{figure}

\begin{figure}[h!]
  \begin{center}
    \includegraphics[width=0.45\textwidth]{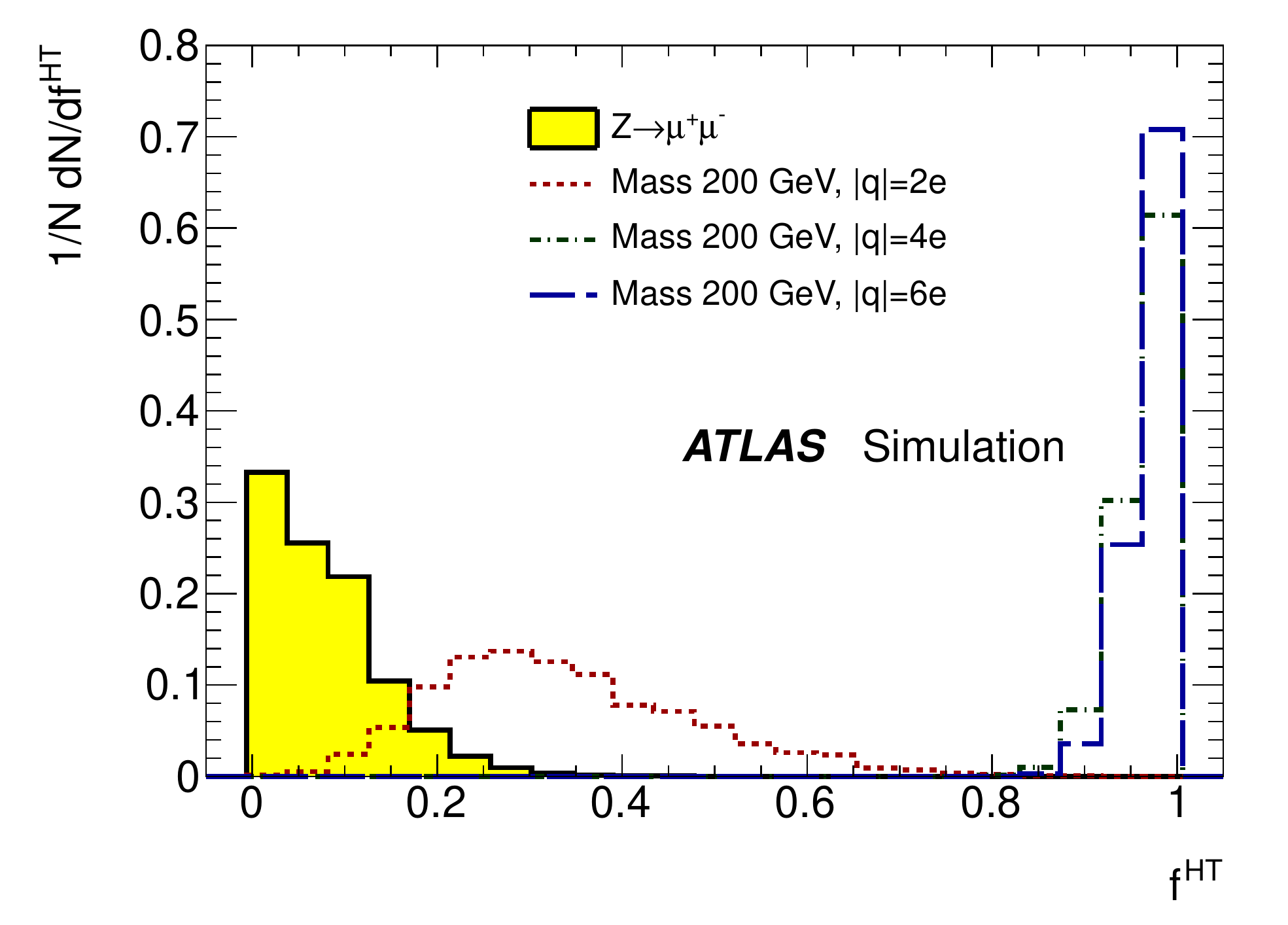}
  \end{center}
  \caption{Normalised distribution of $f^{\textrm{HT}}$ 
for simulated muons and  multi-charged particles. %in arbitrary units. 
Distributions are shown for the signal samples for $|q|$ = $2e$, $4e$, and $6e$ for a mass of \unit[200]{\GeV{}}.
%The signal samples are shown as a function of charge for mass $200$GeV.
}
  \label{fig:TRTfHT_ChargesSeparation}
\end{figure}

An initial preselection of highly ionising candidates is based on the
 pixel \dedx{} significance and the  TRT  high-threshold fraction $f^{\textrm{HT}}$.
As seen in Fig. \ref{fig:pixelseparation}, the pixel \dedx{} significance is a powerful
discriminator for particles with $|q|=2e$. 
The signal region is defined by candidates with a significance greater than \pixSCut{}.
For higher values of $|q|$, the
pixel readout saturates and the \dedx{} signal is no longer reliable. Therefore, to search for particles with $|q|>2e$, the TRT  $f^{\textrm{HT}}$
(see Fig. \ref{fig:TRTfHT_ChargesSeparation}) is used as a discriminating
variable instead. The signal region is defined by requiring the  $f^{\textrm{HT}}$
to be above \htCut{}. This preselection using the pixel \dedx{} or the  $f^{\textrm{HT}}$
 reduces the background contribution by almost three orders of magnitude for both
$|q|$ = $2e$ and $|q| > 2e$.

\begin{figure}[h!]
  \begin{center}
    \includegraphics[width=0.45\textwidth]{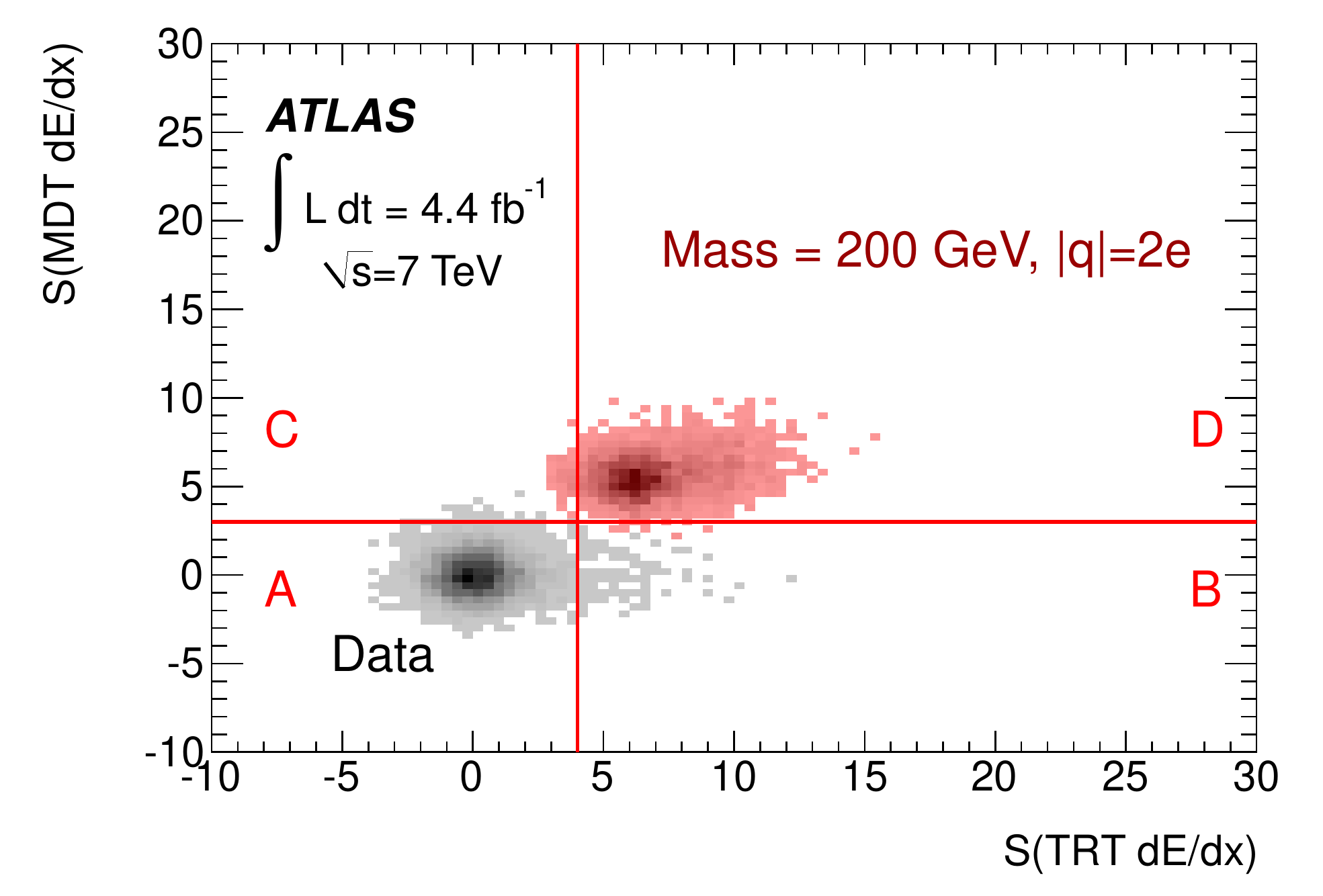}
  \end{center}
  \caption{The plane of TRT and MDT \dedx{}
    significances after the $|q|=2e$ selection. The distributions of the
    2011 data and the signal sample (here for a mass of \unit[200]{\GeV{}}) are shown. 
    The regions labelled A, B and C are control regions used to estimate the background expected in the signal region D.}
  \label{fig:abcplaneq2}
\end{figure}

\begin{figure}[h!]
  \begin{center}
\label{fig:abcplaneq35}
  \includegraphics[width=0.45\textwidth]{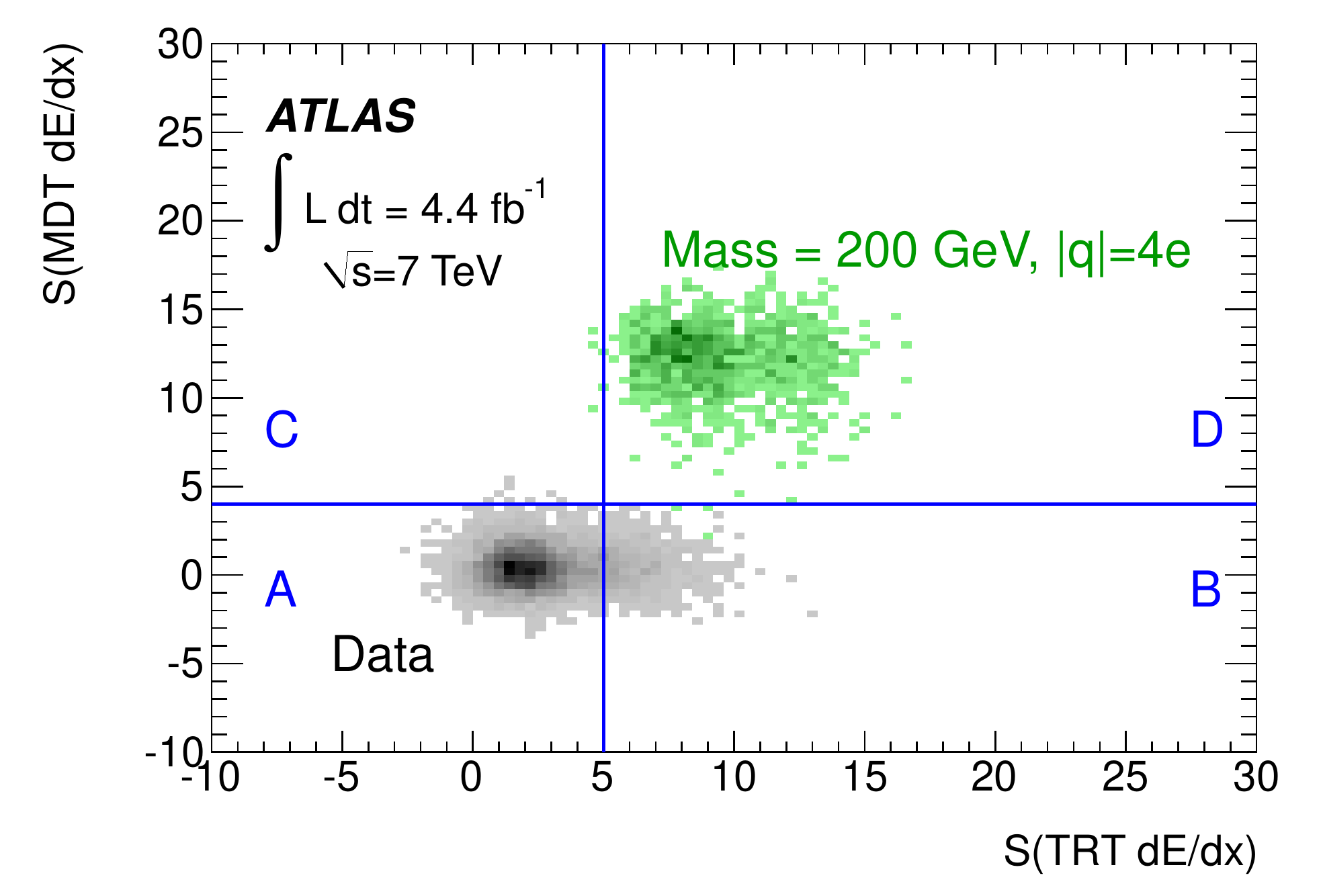}
  \end{center}
  \caption{The plane of TRT and MDT \dedx{}
    significances after the $|q|>2e$ selection. The distributions of the
    2011 data and the signal sample (here for
    a mass of \unit[200]{\GeV{}} and $|q|=4e$) are shown. 
    The regions labelled A, B and C are control regions used to estimate the background expected in the signal region D.}
  \label{fig:abcplaneq3}
\end{figure}

In the final step of the search, the MDT \dedx{} significance, $S$(MDT \dedx{}),
and the TRT \dedx{} significance, $S$(TRT \dedx{}), are used as discriminating 
variables to separate the signal and background. These variables are shown
for real data and simulated signal events in Fig. \ref{fig:abcplaneq2} (\ref{fig:abcplaneq3}) for candidates
preselected as $|q| = 2e$ ($|q| > 2e$). 
Only the signal sample for a mass of \unit[200]{\GeV{}} is shown as there
is very little change in the selection variables for different masses.
As seen, the detector signatures are
different for the two preselected samples, and thus the final signal
regions are chosen differently. They are defined in Table \ref{tab:selection}. 
The selection was optimised using only simulated 
samples and data control samples without examining the signal region in the data.

\begin{table}[htbp]
\caption{The final signal regions for the two preselections.}
\begin{center}
\begin{tabular}{|c||c c|}
\hline
 & $S$(MDT \dedx{}) & $S$(TRT \dedx{})\\
\hline
$|q|=2e$  & $>$ 3 & $>$ 4 \\
$|q|>2e$  & $>$ 4 & $>$ 5  \\
\hline
\end{tabular}
\end{center}
\label{tab:selection}
\end{table}

\section{Background estimation}
\label{sec:abcd}

The background contribution to 
the signal region is estimated using an ABCD method. 
In this method, the regions A,B,C and D are defined by dividing the plane of the uncorrelated TRT and MDT \dedx{} significances 
using the final selection cuts,
as seen in Figs. \ref{fig:abcplaneq2} and \ref{fig:abcplaneq3}. The region D is defined as the signal
region, with regions A, B and C as control regions for the background. 
The expected number of candidates from background in the region D, $\ndata^{\textrm{D}}$, is estimated
from the numbers of observed data candidates in regions A, B and C
($\ndata^{\textrm{A,B,C}}$): 

\begin{equation}
\label{eqn:ABCD}
\ndata^{\textrm{D}} = \frac{\ndata^{\textrm{B}}\times \ndata^{\textrm{C}}}{\ndata^{\textrm{A}}}.
\end{equation}
Table~\ref{tab:yields_bkg} gives the number of candidates in A, B and C, as well as the 
observed number of candidates in the signal region D after the final selection. 
These results are compared to the expected
number of background candidates of \backgroundEstQTwo{} for the $|q| = 2e$ selection
and  \backgroundEstQThree{} for the $|q| > 2e$ selection. The uncertainties are
statistical.  The systematic uncertainty on the background estimation is discussed in Section~\ref{sec:beu}.

\begin{table}[htbp]
\caption{The observed candidate yields in data for an integrated luminosity of
  \intlumi. The last column shows the expected background in the signal region D with 
statistical uncertainty. }
\begin{center}
\begin{tabular}{|c||c c c c|c|}
\hline
 & A & B & C & D & D$_{\textrm{exp.}}$\\
\hline
$|q|=2e$  & 8543 & 92 & 38 &0 & \backgroundEstQTwo{} \\
$|q|>2e$  & 4940 & 754 & 9 &0 &\backgroundEstQThree{} \\
\hline
\end{tabular}
\end{center}
\label{tab:yields_bkg}
\end{table}

\section{Signal selection efficiency}
\label{sec:efficiency}

The signal cross section is given by

\begin{equation}
\sigma = \frac{\nrec}{ 2\times\mathcal{L}\times\epsilon},
\end{equation}
where $\mathcal{L}$ is the integrated luminosity of the analysed data,
$\nrec$ the number of candidate particles in data above the expected background and the factor of 2 is the number of particles per event in the DY model. 
The efficiency  $\epsilon$ includes trigger, reconstruction and selection efficiencies.
The efficiency is the number 
of all multi-charged particles that satisfy the selection criteria
divided by the number of all simulated multi-charged particles.

The efficiency to find a multi-charged particle is given in Table~\ref{tab:overallEff} for each
signal sample.
Several factors contribute to the overall low efficiency and its dependencies on mass and charge. The $|\eta| < 2.0$ selection and  the requirement to reach the MS with a $\beta$ which fits the timing window for the trigger are the primary causes of the reduction in efficiency.  
For the simulated signal samples, this timing requirement generally implies a momentum requirement stricter than the explicit $p_{\text{T}}$ selection.  The implied selection can be as high as approximately $p_{\text{T}}/q >$ \unit[120]{\GeV{}}.
The charge dependence of the efficiency results from higher ionisation and the higher effective single muon $p_{\text{T}}$ selection, which are augmented by the factors $q^2$ and $q$ respectively.  The mass dependence has two competing factors: at low mass there are more candidates above $|\eta| = 2.0$, while at high mass the $\beta$ spectrum is softer.

\begin{table}[!thpb]
 \caption{The efficiencies to select a signal candidate  (in \%)  for the
  DY production model.}
\centering
\begin{tabular} {| c || c| c| c| c| c |}
\hline
Mass & \multicolumn{5}{c |}{Efficiencies [\%]}\\
\cline{2-6}
[GeV] & $|q|=2e$ & $|q|=3e$ & $|q|=4e$ & $|q|=5e$ & $|q|=6e$ \\
\hline
50     & 4.3   &  2.0  &  0.3   &  0.03   &  0.003 \\
100    & 8.6   &  5.5  &  2.3   &  0.4    &  0.07  \\
200    & 12.6  &  9.2  &  4.6   &  1.8    &  0.5   \\
300    & 12.6  &  9.9  &  5.8   &  2.5    &  0.8   \\
400    & 10.9  &  9.0  &  5.6   &  2.9    &  1.0   \\
500    & 9.9   &  8.5  &  5.3   &  2.9    &  1.3   \\
600    & 7.8   &  6.8  &  4.6   &  2.3    &  1.1   \\
\hline
\end{tabular}
\label{tab:overallEff}
\end{table}

\section{Systematic uncertainties}
\label{sec:systematics}

The systematic uncertainties on the background estimate and on the signal efficiency 
are determined by varying %the selection cuts within their uncertainties.
the selection cuts within the uncertainty on each selection variable.

\subsection{Background estimation uncertainty}
\label{sec:beu}
The background estimate in the signal region, D, relies on the fact that the $S$(TRT \dedx{}) and the $S$(MDT \dedx{})  are
 uncorrelated.
To estimate potential influences of signal contamination close to the
region boundaries and remaining correlations in the tails of the
distributions, the ABCD regions are varied. For this estimate, the
signal region D is maintained, but regions A, B and C are redefined by
excluding the region close to the default cut from the background estimation.   This ensures a higher background purity.”
This test is performed for many different
definitions of the control regions and leads to an uncertainty of 5\% on the
estimated background contribution in the signal region.\\

\subsection{Trigger efficiency uncertainty}
The uncertainty on the trigger efficiency has two sources: the standard uncertainty on the trigger efficiency of
1\% as determined by ATLAS muon performance studies \cite{MuonPerformance}
and a $\beta$-dependent trigger uncertainty.
The size of the $\beta$-dependent part is dominated by the uncertainty on the timing correction of the RPC trigger efficiency (trigger for $|\eta| <$ 1.05).  
This correction is varied by $\pm 50\%$ to account for the large dependence of the efficiency on the trigger timing.  
The relative difference of the trigger
efficiencies between the nominal and the varied correction depends on
the mass and charge of the benchmark samples, and ranges from less
than 1\% for $|q|=6e$, $m=\unit[50]{\GeV{}}$ to 24\% for $|q|=5e, m= \unit{600}{\GeV{}}$.   The timing in the TGC (trigger for $|\eta| \ge $ 1.05) for data and simulation is in good agreement, and the systematic uncertainty for the TGC timing correction is negligible. 
The systematic uncertainty on whether a candidate particle would reach
the MS in the timing window for the trigger selection also depends on
the simulation of energy losses in the calorimeters and the material
description of the detector. In a study using muons from $Z\to\mu\mu$
events in data and simulation, the energy losses were shown to be in
excellent agreement.
The energy-loss difference between data and simulation is less than 5\%. A cross-check that
varies the amount of material by $\pm$10\% has a negligible effect on the total systematic uncertainty.

\subsection{Uncertainties due to selection}

The uncertainties on the selection efficiency  arise from the
uncertainties on each selection variable used. The 
following variations of the nominal cuts are studied:
$p_{\text{T}}$ by $\pm3$\%, $S$(pixel \dedx{}) by $\pm5$\%, TRT HT fraction by $\pm20$\%,
$S$(TRT \dedx{}) by $\pm5$\% and $S$(MDT \dedx{}) by $-5$\% and $+50$\%.
For the $p_{\text{T}}$ cut this corresponds to the resolution of the track
$p_{\text{T}}$ measurements. The variation of 20\% of the TRT HT fraction
arises from the pile-up dependence of this variable. For
the  pixel and the TRT \dedx{} significances, 5\% corresponds to the observed
agreement of the mean and width of these distributions in the $Z\to\mu\mu$ events in data and simulation. This is also applied to  the lower
variation of $S$(MDT \dedx{}).  
Here, a relative shift between simulation and data is observed.
The magnitude and direction of this shift suggest
a variation of $S$(MDT \dedx{}) by 50\% in the positive direction.  
While this would have been important for a potential signal
interpretation, it has only a small effect on the limit setting. 
For all other variables the variations have no observable effect in any
of the signal samples. The total systematic uncertainties on the efficiency arising from these
cut variations range up to 2.1\%.

\subsection{Summary of systematic uncertainties}
In Table~\ref{tab:syst_source_sum_ind} the quadratic sums of all
the systematic uncertainties considered above are summarised for the different signal
samples. The two main uncertainties are the uncertainty on
the trigger efficiency and the uncertainty due to the
small number of Monte Carlo events. The latter makes a significant contribution for some of the high-charge and
low-mass samples.  The \unit[50]{\GeV{}} samples were produced with a selection
at the generator level requiring $p_{\text{T}}/q >$ \unit[15]{\GeV{}} in order to decrease the statistical uncertainty.
The systematic uncertainties vary between 6\% and 28\% in total.

\begin{table}[h]
\caption{Summary of relative systematic uncertainties on the expected number of candidates
  derived from the uncertainties on the background estimation,
   trigger efficiency, Monte Carlo statistics and due to selection cuts.} 
\begin{center}
\begin{tabular}{|c||c|c|c|c|c|}
\hline
Mass & \multicolumn{5}{|c|}{quadratic sum of systematic uncertainties [\%]}\\
\cline{2-6}
[GeV] & $|q|$ = $2e$ &$|q|$ = $3e$&$|q|$ = $4e$&$|q|$ = $5e$&$|q|$ = $6e$\\
\hline
50     &  8    &  6    &   6   &  10   &  19   \\
100    &  10   & $9  $ & $ 7 $ & $12 $ & $28 $ \\
200    & $13 $ & $12 $ & $10 $ & $ 9 $ & $12 $ \\
300    & $14 $ & $15 $ & $15 $ & $12 $ & $11 $ \\
400    & $17 $ & $17 $ & $18 $ & $18 $ & $13 $ \\
500    & $18 $ & $18 $ & $19 $ & $21 $ & $18 $ \\
600    & $22 $ & $22 $ & $23 $ & $25 $ & $24 $ \\
\hline
\end{tabular}
\end{center}
\label{tab:syst_source_sum_ind}
\end{table}

The uncertainty on the integrated luminosity is estimated to be 3.9\% from Van der Meer
scans~\cite{Aad:2011dr,ATLAS-CONF-2011-116} and is not included in Table~\ref{tab:syst_source_sum_ind}.

\section{Results}
\label{sec:limits}

\begin{figure*}[!thpb]
 \centering
 \includegraphics[width=0.98\textwidth]{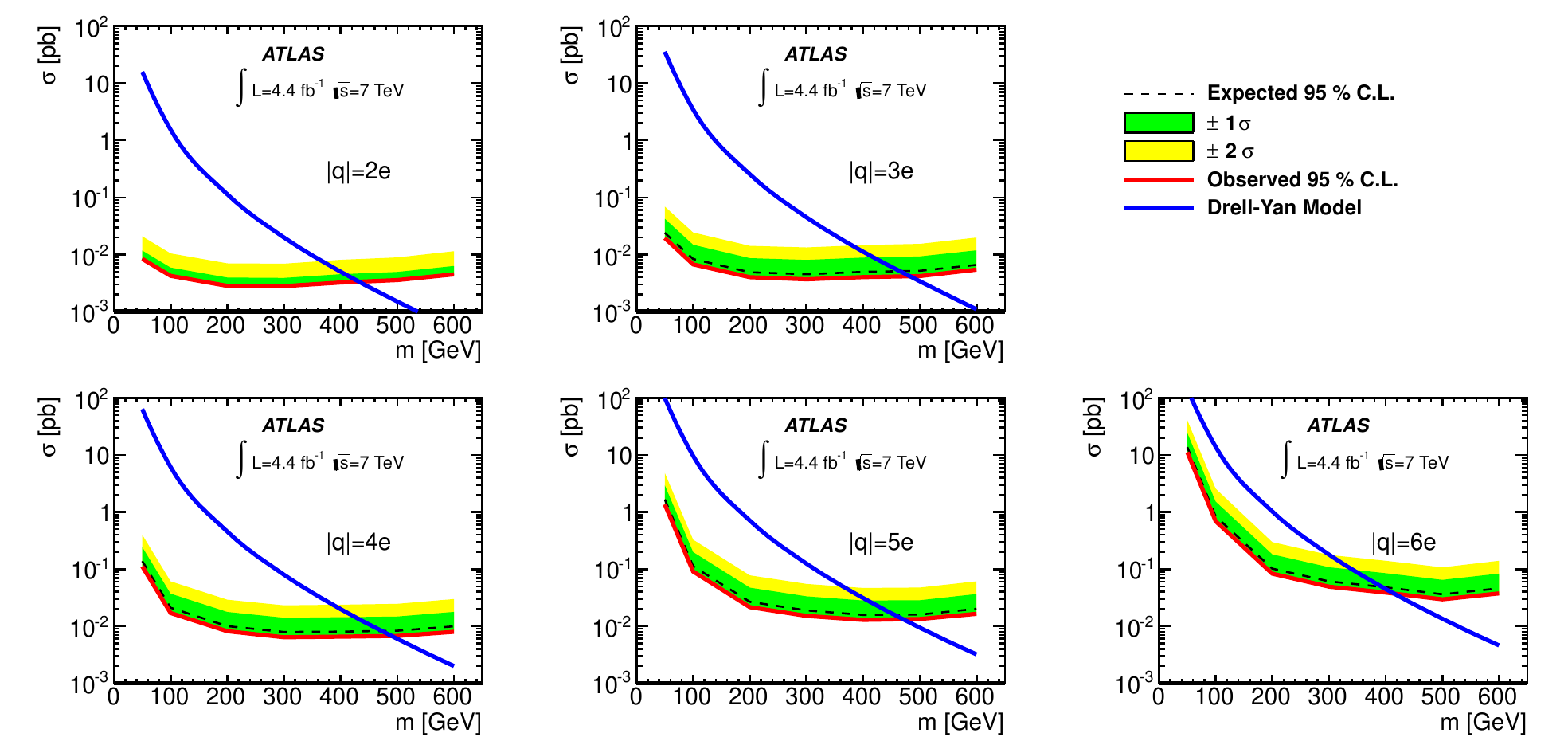}
 \caption{Upper limits on the production cross section of
%95\% CL limits on the production cross section of
   multi-charged highly ionising particles from pair-production
   as a function of particle mass. The dotted line shows the
   expected limit with the $\pm1\sigma$ and $\pm2\sigma$
 uncertainty bands. The observed limit is compared with the
 predicted rapidly falling cross section from the DY model. The plots
 are shown separately for charges from $|q|=2e$ to $|q|=6e$. In the $|q|=2e$ case, the observed limit lies on top of the expected limit.}
 \label{fig:limitseparate}
 \end{figure*}

\begin{figure}[p]
 \centering
 \includegraphics[width=0.45\textwidth]{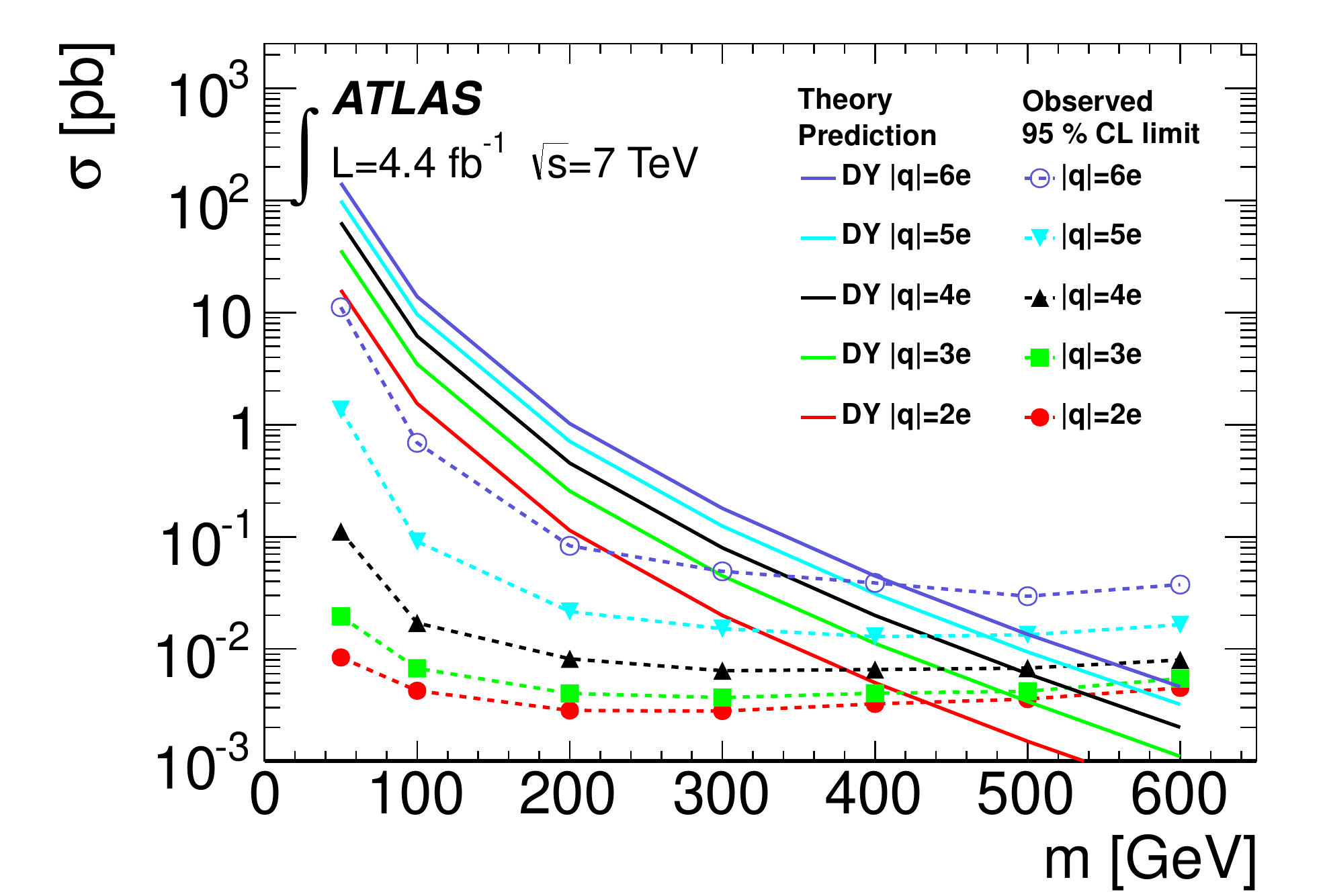}
 \caption{Observed 95\% CL cross-section upper limits and theoretical cross sections as functions of the multi-charged particle mass. }
 \label{fig:limtrack}
 \end{figure}

No signal candidates are found for either the $|q|=2e$ or the $|q|>2e$ selected sample. The results are consistent
with the expectation of \backgroundEstQTwoStatSys{} and
\backgroundEstQThreeStatSys{} background candidates, %for the $|q|=2e$ and $|q|>2e$ split,
respectively. From these numbers the expected and observed limits are
computed using pseudo-experiments.
For the total cross-section limit, the systematic uncertainties on
efficiency and the luminosity are taken into account in the
pseudo-experiments. 
For every benchmark point, 100,000 pseudo-experiments 
are used. The measurement
excludes DY model pair-production over wide ranges of tested masses. 
Fig.~\ref{fig:limitseparate} shows the observed 95\% confidence level cross-section limits
as a function of mass for  the five different charges. 
Due to the low number of expected events, the dominant uncertainty arises from  Poisson statistics as reflected in the asymmetric uncertainty bands.
The limits range from around $10^{-2}$ pb for the lower charges to $10^{-1}$ pb for $|q| = 6e$.  In addition to the expected and
observed limits the predicted  cross section is shown for the
simplified Drell--Yan model. %used for the efficiency calculation.  
For the given model the cross-section limits can be transformed into mass
exclusion lower limits from \unit[50]{\GeV{}} to 430, 480, 490, 470 and \unit[420]{GeV} for charges $|q| = 2e$, $3e$,
$4e$, $5e$ and $6e$, respectively. Fig.~\ref{fig:limtrack} summarises the observed
limits. 

\section{Summary}
\label{sec:summary}
A search for long-lived, multi-charged
particles has been performed using an integrated luminosity of \intlumi 
of $pp$ collisions recorded by the ATLAS
detector at the LHC. No candidates are found in the 2011 data
set,  consistent with the background expectation.
The results presented here are the first mass limits from ATLAS for charges of $2e$
to $6e$,  filling the missing range of charges between the searches for slow singly charged long-lived
particles \cite{SMP2012} and searches for particles with charges from $6e$ to $17e$ \cite{HIP2010}.

% Acknowledgements for papers with collision data
% Version 5-Nov-2012

\section{Acknowledgements}

% Standard acknowledgements start here
%----------------------------------------------
We thank CERN for the very successful operation of the LHC, as well as the
support staff from our institutions without whom ATLAS could not be
operated efficiently.

We acknowledge the support of ANPCyT, Argentina; YerPhI, Armenia; ARC,
Australia; BMWF and FWF, Austria; ANAS, Azerbaijan; SSTC, Belarus; CNPq and FAPESP,
Brazil; NSERC, NRC and CFI, Canada; CERN; CONICYT, Chile; CAS, MOST and NSFC,
China; COLCIENCIAS, Colombia; MSMT CR, MPO CR and VSC CR, Czech Republic;
DNRF, DNSRC and Lundbeck Foundation, Denmark; EPLANET, ERC and NSRF, European Union;
IN2P3-CNRS, CEA-DSM/IRFU, France; GNSF, Georgia; BMBF, DFG, HGF, MPG and AvH
Foundation, Germany; GSRT and NSRF, Greece; ISF, MINERVA, GIF, DIP and Benoziyo Center,
Israel; INFN, Italy; MEXT and JSPS, Japan; CNRST, Morocco; FOM and NWO,
Netherlands; BRF and RCN, Norway; MNiSW, Poland; GRICES and FCT, Portugal; MERYS
(MECTS), Romania; MES of Russia and ROSATOM, Russian Federation; JINR; MSTD,
Serbia; MSSR, Slovakia; ARRS and MVZT, Slovenia; DST/NRF, South Africa;
MICINN, Spain; SRC and Wallenberg Foundation, Sweden; SER, SNSF and Cantons of
Bern and Geneva, Switzerland; NSC, Taiwan; TAEK, Turkey; STFC, the Royal
Society and Leverhulme Trust, United Kingdom; DOE and NSF, United States of
America.

The crucial computing support from all WLCG partners is acknowledged
gratefully, in particular from CERN and the ATLAS Tier-1 facilities at
TRIUMF (Canada), NDGF (Denmark, Norway, Sweden), CC-IN2P3 (France),
KIT/GridKA (Germany), INFN-CNAF (Italy), NL-T1 (Netherlands), PIC (Spain),
ASGC (Taiwan), RAL (UK) and BNL (USA) and in the Tier-2 facilities
worldwide.
%----------------------------------------------

\providecommand{\href}[2]{#2}\begingroup\raggedright

\endgroup

\onecolumn
\clearpage
% ATLAS Collaboration author list
% Data extracted on 29-Jan-2013 for paper reference EXOT-2012-10
%\documentclass[11pt]{article}
%\usepackage{a4wide}\begin{document}
\begin{flushleft}
{\Large The ATLAS Collaboration}

\bigskip

G.~Aad$^{\rm 48}$,
T.~Abajyan$^{\rm 21}$,
B.~Abbott$^{\rm 111}$,
J.~Abdallah$^{\rm 12}$,
S.~Abdel~Khalek$^{\rm 115}$,
A.A.~Abdelalim$^{\rm 49}$,
O.~Abdinov$^{\rm 11}$,
R.~Aben$^{\rm 105}$,
B.~Abi$^{\rm 112}$,
M.~Abolins$^{\rm 88}$
% ...
\bigskip

% ...
$^{11}$ Institute of Physics, Azerbaijan Academy of Sciences, Baku, Azerbaijan\\
$^{12}$ Institut de F\'{i}sica d'Altes Energies and Departament de F\'{i}sica de la Universitat Aut\`{o}noma de Barcelona and ICREA, Barcelona, Spain\\
% ...
$^{21}$ Physikalisches Institut, University of Bonn, Bonn, Germany\\
% ...
$^{48}$ Fakult\"{a}t f\"{u}r Mathematik und Physik, Albert-Ludwigs-Universit\"{a}t, Freiburg, Germany\\
$^{49}$ Section de Physique, Universit\'{e} de Gen\`{e}ve, Geneva, Switzerland\\
% ...
$^{88}$ Department of Physics and Astronomy, Michigan State University, East Lansing MI, United States of America\\
% ...
$^{105}$ Nikhef National Institute for Subatomic Physics and University of Amsterdam, Amsterdam, Netherlands\\
% ...
$^{111}$ Homer L. Dodge Department of Physics and Astronomy, University of Oklahoma, Norman OK, United States of America\\
$^{112}$ Department of Physics, Oklahoma State University, Stillwater OK, United States of America\\
% ...
$^{115}$ LAL, Universit\'{e} Paris-Sud and CNRS/IN2P3, Orsay, France\\
% ...
\end{flushleft}

%\end{document}
% Created with TexConverter

\end{document}